\documentclass[usenatbib,fleqn]{mn2e}

\usepackage[american]{babel}
\usepackage{indentfirst}
\usepackage{mathrsfs}
\usepackage{graphicx}
\usepackage{color}
\usepackage{amsmath}
\usepackage{amssymb}
\usepackage{amsfonts}

\newcommand{\de}{\mathrm d}

\title{Weak lensing signal in Unified Dark Matter models}
\author[S. Camera et al.]{S. Camera$^{1,2,\star}$, D. Bertacca$^{3,4,\dagger}$, A. Diaferio$^{1,2,\ddagger}$, N. Bartolo$^{3,4,\sharp}$ \& S. Matarrese$^{3,4,\flat}$\\$^1$Dipartimento di Fisica Generale ``Amedeo Avogadro'', Universit\`a degli Studi di Torino, Torino, Italy\\$^2$Istituto Nazionale di Fisica Nucleare (INFN), Sezione di Torino, Torino, Italy\\$^3$Dipartimento di Fisica ``Galileo Galilei'', Universit\`a di Padova, Padova, Italy\\$^4$Istituto Nazionale di Fisica Nucleare (INFN), Sezione di Padova, Padova, Italy\\\newline\\$^\star$camera@ph.unito.it\\$^\dagger$bertacca@pd.infn.it\\$^\ddagger$diaferio@ph.unito.it\\$^\sharp$bartolo@pd.infn.it\\$^\flat$matarrese@pd.infn.it}
\date{}

\begin{document}

\maketitle

\begin{abstract}

Weak gravitational lensing is a powerful tool for studying both the geometry and the dynamics of the Universe. Its power spectrum contains information on the sources emitting photons and on the large--scale structures that these photons cross. We calculate the weak lensing cosmic convergence and shear power spectra, in linear theory and Limber's approximation, for two different classes of cosmological models: the standard $\Lambda$CDM and Unified Dark Matter (UDM) models. The latter models attempt to solve the problems of the dark matter in the dynamics of galaxies and galaxy clusters and of the late--time acceleration of the Universe expansion by introducing a scalar field that mimics both dark matter and dark energy. A crucial feature of the UDM models is the speed of sound $c_\infty$, that is the value of the sound speed at late times, on which structure formation depends. In this paper, we provide the predictions of the UDM models for the weak lensing signal,with various values of $c_\infty$. We consider both the Cosmic Microwave Background and background galaxies at different redshifts as sources for the weak lensing power spectra. We find that the power spectra in UDM models are more sensitive to the variations of $c_\infty$ for sources located at low redshifts. Moreover, we find that for $c_\infty>10^{-3}$ (in units of the speed of light), the UDM weak lensing convergence power spectrum $C^{\kappa\kappa}(l)$ for background galaxies is strongly suppressed with respect to the $\Lambda$CDM spectrum, particularly at small angular scales $l\gtrsim100$.
\end{abstract}


\section{Introduction}

The Standard Cosmological Model is based on two cornerstones: General Relativity (GR) and the Standard Model of Fundamental Interactions (SM). While the latter describes the content of the Universe in terms of matter and energy, the former involves only the gravitational interaction. In the last decades cosmologists have obtained a large amount of observational data and could constrain their theory. However, a number of independent data sets, the distance modulus from Supernov\ae~Ia used as standard candles \citep{Riess:1998cb,Knop:2003iy,Riess:2004n,Riess:2006fw}, the Cosmic Microwave Background (CMB) anisotropies \citep{deBernardis:2000gy,Bennett:2003bz,Spergel:2003cb,Hinshaw:2008kr,Komatsu:2008hk} and the Large-Scale Structure (LSS) \citep{Zwicky:1933gu,Zwicky:1937zza,Dodelson:2001ux,Hawkins:2002sg,Spergel:2006hy} have now firmly established that the Universe is undergoing a period of accelerated expansion and moreover that baryonic matter is insufficient to explain the observed dynamics of the LSS. There are two approaches to explain these two pieces of observational evidence: either a modification to SM or to GR. The most promising solution seems to be the former, and in fact cosmologists add new forms of matter and energy in the model. Data from LSS and CMB agree with the dynamics dominated by Dark Matter (DM). The cosmic acceleration is explained by the presence of a cosmological constant $\Lambda$, or a more general Dark Energy (DE) component. However, DM is a cold collisionless component mostly made of hypothetical elementary particles (e.g. Weakly Interacting Massive Particles) that have not yet been detected; and $\Lambda$ appears to be so tiny that there is no theoretical justification for it \citep{1989RvMP...61....1W}.

In this paper, we focus on unified models of DM and DE (UDM), in which a single scalar field provides an alternative interpretation to the nature of the dark components of the Universe. Compared with the standard DM + DE models (e.g. $\Lambda$CDM), these models have the advantage that we can describe the dynamics of the Universe with a single dark fluid which triggers both the accelerated expansion at late times and the LSS formation at earlier times. While most of the models of DE relies on the potential energy of scalar fields to lead to the late time acceleration of the Universe, it is possible to have a situation where the accelerated expansion arises out of modifications to the kinetic energy of the scalar field. Originally this method was proposed to have kinetic energy driven inflation, called $k-$inflation \citep{ArmendarizPicon:1999rj}, to explain early Universe inflation at high energies. Then this scenario was applied to DE \citep{Chiba:1999ka,Linder:2008ya,dePutter:2007ny}. The analysis was extended to a more general Lagrangian \citep{ArmendarizPicon:2000dh,ArmendarizPicon:2000ah} and this scenario was called $k-$essence.  Among all the models, there are two families of adiabatic $k-$essence \citep{Bertacca:2007ux}: $i)$ the generalized Chaplygin gas \citep{Kamenshchik:2001cp,Bilic:2001cg,Bento:2002ps,Sandvik:2002jz,Carturan:2002si} and the single dark perfect fluid with a simple two-parameter barotropic equation of state \citep{Balbi:2007mz,Quercellini:2007ht,Pietrobon:2008js}; $ii)$ the purely kinetic models \citep{Scherrer:2004au,Bertacca:2007ux}. The major advantage of all these models is that there is only one
non-standard fluid, which can mimic both DM and DE. Thus, one of the main issues of these UDM models is to see whether the single dark fluid is able to cluster and produce the cosmic structures we observe in the Universe today. In fact, the effective speed of sound can be significantly different from zero at late times; the corresponding Jeans' length (or sound horizon), below which the dark fluid cannot cluster, can be so large that the gravitational potential first strongly oscillates and then decays \citep{Bertacca:2007cv,Hu:1998kj}, thus preventing structure formation. Here we choose to investigate the class of scalar field Lagrangians with a non-canonical kinetic term that
allow to obtain UDM models with a small effective sound speed.

Recently, it has been shown that the scalar field in UDM models can cluster \citep{Bertacca:2008uf}, but it remains to be explored whether UDM models provide a good fit to the various sets of available data. Weak lensing is a powerful tool \citep{Hu:2000ee,Hu:2001fb,Hu:1998az,Schmidt:2008hc,Thomas:2008tp}; gravitational lens effects are due to the deflection of light occuring when photons travel near matter, i.e. in the presence of a non-neglegible gravitational field. The cosmic convergence and shear encapsulate information about both the source emitting light and the structures that photons cross before arriving at the telescope. Hence, weak lensing allows to explore both the basis of the cosmological model and LSS of the Universe, in other words it brings information about the geometry and the dynamics. Therefore the study of the power spectrum of weak lensing can be a crucial test.

In this paper we focus on weak gravitational lensing in UDM models, for CMB and galaxy photons. In \S\ref{cosmologicalmodels}, $\Lambda$CDM and UDM models are briefly presented. In \S\ref{weaklensing}, the standard framework of weak gravitational lensing phenomena is presented. In \S \ref{powerspectra}, power spectra of weak lensing, convergence and shear are computed, on both the all sky and in flat--sky approximation (Limber's approximation, \citealt{Kaiser:1991qi}). In \S \ref{results} we present the results for the convergence power spectrum for CMB photons (\S \ref{cmb}) and light from background galaxies (\S \ref{backgroundgalaxies}). In \S \ref{conclusions} conclusions are drawn.

\section{Cosmological models}\label{cosmologicalmodels}

The metric of the Universe is described by the standard Friedmann--Lema\^itre--Robertson--Walker (FLRW) length element for an isotropic and homogeneous space filled with a perfect fluid. With the addition of scalar perturbations and in Newtonian gauge, the length element takes the form\footnote{We use units such that $c=1$ and signature $\{-,+,+,+\}$, where Greek indeces run over spacetime dimension, whereas Latin indeces label spatial coordinates.}
\begin{equation}
\de s^2\equiv g_{\mu\nu}\de x^\mu \de x^\nu=a^2(\tau)\left[-(1+2\mathbf{\Phi})\de\tau^2+(1+2\Psi)\de\ell^2\right]\label{flrw}
\end{equation}
where $\de \tau=\de t/a$ is the conformal time, with the flat spatial metric \citep{Spergel:2006hy}
\begin{equation}
d\ell^2=d\chi^2+\chi^2\left[d\vartheta^2+\sin^2\vartheta\,d\varphi^2\right]
\end{equation}
with the radial comoving distance
\begin{equation}
\chi(z)=r_H\int_0^z\!\!d\tilde z\,\frac{H_0}{H(\tilde z)},
\end{equation}
where $H=\dot a/a$ is the Hubble parameter and $r_H=c/H_0\simeq3\,h^{-1}\,\mathrm{Gpc}$ is the Hubble radius.

\subsection{$\Lambda$CDM}

In $\Lambda$CDM, gravity is described by GR in a 4-dimensional Universe filled with a perfect fluid of photons, baryons, cold DM and DE (as cosmological constant) with stress--energy tensor
\begin{equation}
T_{\mu\nu}=(p+\rho)u_\mu u_\nu-p\,g_{\mu\nu}\label{stress--energy}.
\end{equation}
where $p(t,\mathbf x)$ is the pressure, $\rho(t,\mathbf x)$ is the energy density and $u_\mu$ is the quadrivelocity. In order to simplify further the calculations, from now on we will consider no longer the contributions of baryons.

Since the anisotropic stress is negligible, $T_{ij}=0$ for $i\neq j$, and the only solution of linearized Einstein's equations for \eqref{flrw} and \eqref{stress--energy} consistent with $\mathbf{\Phi}$ and $\Psi$ being perturbations is $\mathbf{\Phi}\equiv-\Psi$. For scalar adiabatic perturbations, introducing a speed of sound ${c_s}^2=\partial p/\partial\rho$, the complete set of Einstein's equations can be reduced to \citep{Mukhanov:2005sc}
\begin{equation}
v''-{c_s}^2\nabla^2v-\frac{\theta''}{\theta}v=0\label{eq-Mukhanov:2005sc-lcdm}
\end{equation}
where a prime denote a derivative with respect to the conformal time,
\begin{align}
v&=\frac{\mathbf{\Phi}}{\sqrt{\rho+p}}\label{udiphi-lcdm},\\
\theta&=\frac{1}{a}\left(1+\frac{p}{\rho}\right)^{-\frac{1}{2}}.
\end{align}
We are dealing with the gravitational potential after recombination, and so there is no more sound speed due to radiation. Thus, Eq.~\eqref{eq-Mukhanov:2005sc-lcdm} has analytic solution in Fourier's space \citep{Hu:1998tj,Hu:2001fb,Mukhanov:2005sc,Bertacca:2007cv}
\begin{equation}
\mathbf{\Phi}_k(a)=A_k\left(1-\frac{H(a)}{a}\int_0^a\!\!\frac{d\tilde a}{H(\tilde a)}\right)
\end{equation}
where the constant of integration is $A_k=\mathbf{\Phi}_k(0)T(k)$, with $T(k)$ the matter transfer function, that describes the evolution of perturbations through the epochs of horizon crossing and radiaton-matter transition, and $\mathbf{\Phi}_k(0)$ the large-scale potential during the radiation dominated era. We emphasize that this approach to perturbation theory is completely equivalent to the standard one \citep[e.g.][]{Dodelson:2003ft}.

The Hubble parameter as a function of redshift $z$ is given by
\begin{equation}
H(z)=H_0\sqrt{\Omega_m{(1+z)}^3+\Omega_\Lambda}\label{HLCDM}.
\end{equation}
with $H_0=100\,h\,\mathrm{km\,s^{-1}\,Mpc^{-1}}$ the Hubble parameter today, $\Omega_m$ the matter density and $\Omega_\Lambda$ the contribution of the cosmological constant \citep{Lahav:1991wc}.

\subsection{UDM}
\label{UDMmodels}

UDM models use a scalar field $\varphi(t,\mathbf x)$ that mimics both DM and DE. This can be achieved with a non-canonical kinetic term, i.e. by letting the kinetic energy be a generic function of the derivatives of the scalar field \citep{Bertacca:2007ux}, instead of the standard form $\dot\varphi^2/2$. The Lagrangian density can be written as
\begin{equation}
\mathscr L= \mathscr L_G+\mathscr L_\varphi=\frac{1}{16\pi G}R+\mathscr L_\varphi(\varphi,X)
\end{equation}
where
\begin{equation}
X=-\frac{1}{2}\nabla_\mu\varphi\nabla^\mu\varphi.
\end{equation}
The energy-momentum tensor of the scalar field is defined in the usual way
\begin{equation}
T_{\mu\nu}=-\frac{2}{\sqrt{-g}}\frac{\delta(\mathscr L_\varphi\sqrt{-g})}{\delta g^{\mu\nu}}=\frac{\partial\mathscr L_\varphi}{\partial X}\nabla_\mu\varphi\nabla_\nu\varphi+\mathscr L_\varphi g_{\mu\nu}.
\end{equation}
If $X$ is time-like, $\mathscr L_\varphi$ describes a perfect fluid with pressure
\begin{equation}
p=\mathscr L_\varphi(\varphi,X)
\end{equation}
and energy density
\begin{equation}
\rho=2X\frac{\partial p}{\partial X}-p.
\end{equation}
In \citet{Bertacca:2007ux} the scalar field Lagrangian is required to be constant along the classical trajectories. Specifically, by requiring that $\mathscr L_\varphi\equiv p=-\Lambda/(8\pi G)$ on cosmological scales, the background is identical to the background of $\Lambda$CDM. In the following we will focus on such UDM models. In fact, if we
consider the equation of motion of the scalar field, and
we impose that $p=-\Lambda/(8\pi G)$, we easily get
\begin{equation}
\rho\left[a(t)\right]=\rho_\mathrm{DM}(a=1)a^{-3}+\frac{\Lambda}{8\pi G}\equiv\rho_\mathrm{DM}+\rho_\Lambda\, ,
\end{equation}
where $\rho_\Lambda$ behaves like a cosmological constant DE component ($\rho_\Lambda=\mathrm{const.}$) and $\rho_\mathrm{DM}$ behaves like a DM component ($\rho_\mathrm{DM}\propto a^{-3}$). This result implies that we can think the stress tensor of the scalar field as being made of two components: one behaving like a pressureless fluid, and the other having negative pressure. Therefore the integration constant $\rho_\mathrm{DM}(a = 1)$ can be interpreted as the ``dark matter'' component today; consequently, $\Omega_m=8\pi G\rho_\mathrm{DM}(a=1)/(3{H_0}^2)$ and $\Omega_\Lambda=8\pi G\rho_\Lambda/(3{H_0}^2)$ are the density parameters of DM and DE today, and so the Hubble parameter in UDM is the same as in $\Lambda$CDM, Eq.~\eqref{HLCDM}.

Now we introduce small inhomogeneities of the scalar field $\varphi(t,\mathbf x)$, and we use the Newtonian gauge \eqref{flrw}. We want to stress that gravity is GR, and the scalar field $\varphi(t,\mathbf x)$ presents no anisotropic stress, and so, again, $\mathbf{\Phi}\equiv-\Psi$.
By linearizing Einstein's equations we obtain again the second order differential equation \eqref{eq-Mukhanov:2005sc-lcdm} \citep{Garriga:1999vw,Mukhanov:2005sc}, but, here, $c_s$ is the ``speed of sound'' relative to the pressure and energy density fluctuations of the scalar field, and it is defined by
\begin{equation}
{c_s}^2=\frac{\partial p/\partial X}{\partial\rho/\partial X}=\frac{p_{,X}}{p_{,X}+2Xp_{,XX}},
\end{equation}
where $_{,X}$ denotes a derivative with respect to $X$. It accounts for the presence of intrinsic entropy perturbations of the fluid \citep{Hu:1998kj,Hu:1998tj}. The variables $v$ and $\theta$ coincide with the corresponding quantities of $\Lambda$CDM. But while in $\Lambda$CDM the Universe contains a hydrodynamical fluid, now $v$ and $\theta$ describe the perturbations in the homogeneous scalar condensate.

\begin{figure}
\centering
\input{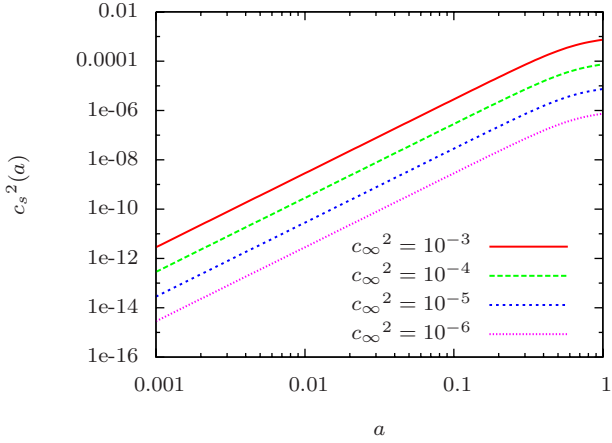}
\caption{Sound speed ${c_s}^2(a)$ for different values of ${c_\infty}^2=10^{-6},10^{-5},10^{-4},10^{-3}$ from bottom to top.}\label{c_s}
\end{figure}

Considering a Lagrangian of the form
\begin{equation}
\mathscr L_\varphi=f(\varphi)g(X)-V(\varphi)
\end{equation}
\citet{Bertacca:2008uf} proposed a technique to construct UDM models where the scalar field can have a sound speed small enough to allow structure formation and to avoid a strong integrated Sachs-Wolfe effect in the CMB anisotropies which tipically plague UDM models. The parametric form for the sound speed is \citep{Bertacca:2008uf}
\begin{equation}
{c_s}^2(a)=\frac{{\Omega_\Lambda c_\infty}^2}{\Omega_\Lambda+(1-{c_\infty}^2)\Omega_ma^{-3}}
\end{equation}
where $c_\infty$ is the value of the sound speed when $a\rightarrow\infty$. Let us emphasize that when $a\to0$, $c_s\to0$. In Fig.~\ref{c_s} we present ${c_s}^2(a)$ for different values of $c_\infty$.

\begin{figure}
\centering
\input{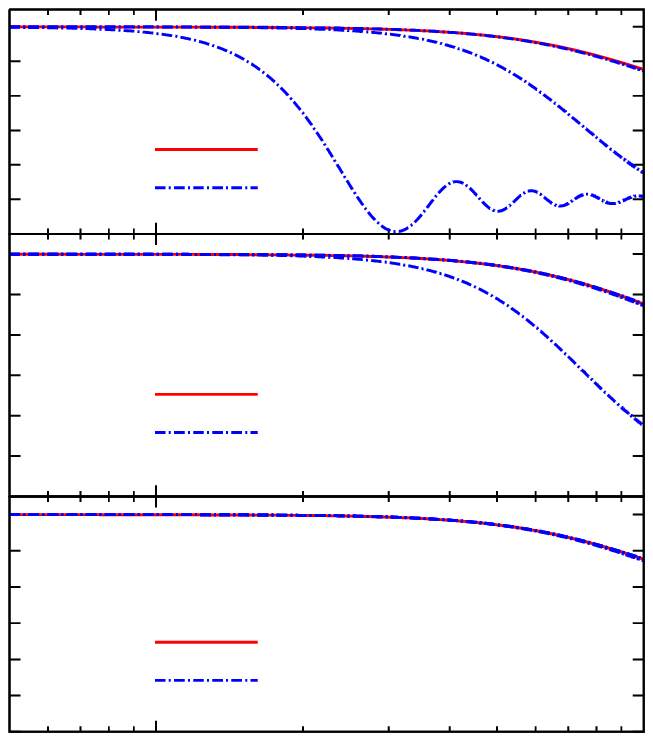}
\caption{Normalized potentials $\mathbf{\Phi}_k(a)/\mathbf{\Phi}_k(0)$ are shown for $\Lambda$CDM (solid) and UDM (dot-dashed). The lower panel shows potentials at $k=0.001\,h\,\mathrm{Mpc}^{-1}$, the medium panel at $k=0.01\,h\,\mathrm{Mpc}^{-1}$ and the upper panel at $k=0.1\,h\,\mathrm{Mpc}^{-1}$. UDM curves are for ${c_\infty}^2=10^{-6},10^{-4},10^{-2}$ from top to bottom, respectively. At small $c_\infty$, $\Lambda$CDM and UDM curves are indistinguishable.}\label{potentials_k-cinf}
\end{figure}
\begin{figure}
\centering
\input{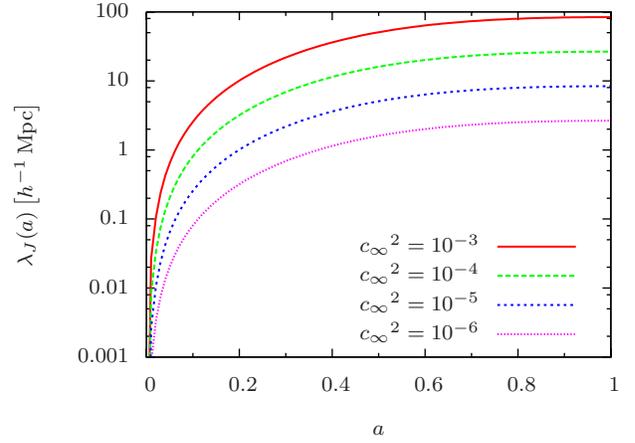}
\caption{Sound horizon $\lambda_J(a)$ for different values of ${c_\infty}^2=10^{-6},10^{-5},10^{-4},10^{-3}$ from bottom to top.}\label{lambda_J}
\end{figure}

In Fig.~\ref{potentials_k-cinf} we present some Fourier's components $\mathbf{\Phi}_k(a)$ of the gravitational potential, normalized to unity at early times. In the case of the UDM models there are two simple but important aspects: first, the fluid which triggers the accelerated expansion at late times is also the one which has to cluster in order to produce the structures we see today. Second, from the last scattering to the present epoch, the energy density of the Universe is dominated by a single dark fluid, and therefore the gravitational potential evolution is determined by the background and perturbation evolution of this fluid alone. As a result, the general trend is that the possible appearance of a sound speed significantly different from zero at late times corresponds to the appearance of a Jeans' length under which the dark fluid does not cluster any more, causing a strong evolution in time of the gravitational potential. By increasing the sound speed, the potential starts to decay earlier in time, oscillating around zero afterwards. Moreover at small scales, if the sound speed is small enough, UDM reproduces $\Lambda$CDM. This reflects the dependence of the gravitational potential on the effective Jeans' length ${\lambda_J}^2(\tau)={c_s}^2|\theta/\theta''|$ \citep{Bertacca:2007cv}. In Fig.~\ref{lambda_J} we show $\lambda_J(a)$, the sound horizon, for different values of $c_\infty$.

\section{Weak gravitational lensing}
\subsection{Weak lensing observables}\label{weaklensing}

From GR we know that light beam paths are curved by the presence of matter. In the weak lensing framework the deflection of light is small and, consequently, we can use Born's approximation, where lensing effects are evaluated on the null-geodesic of the unperturbed (unlensed) photon \citep{Hu:2000ee,Bartelmann:1999yn}.

All weak lensing observables may be expressed in terms of the projected potential
\begin{equation}
\phi(\hat{\mathbf n})=\int\!\!\de\chi\,\frac{W(\chi)}{\chi^2}\mathbf{\Phi}(\hat{\mathbf n},\chi)\label{phi}
\end{equation}
where
\begin{equation}
W(\chi)=-2\chi\int_\chi^\infty\de\chi'\,\frac{\chi'-\chi}{\chi'}n(\chi')\label{W(z)}
\end{equation}
is the weight function of weak lensing, with $n\left[\chi(z)\right]$ representing the redshift distribution of sources, such that $\int\!\!\de\chi\,n(\chi)=1$. We now introduce a distortion tensor
\begin{equation}
\phi_{,ij}(\hat{\mathbf n})\equiv\frac{\partial{\vartheta_s}^i}{\partial\vartheta^j}-\delta_{ij}=\int\!\!\de\chi\,W(\chi)\mathbf{\Phi}_{,ij}(\hat{\mathbf n},\chi)\label{phi,ij}
\end{equation}
${\vartheta_s}^i$ being the value of the displacement at the source image and $\vartheta^i$ the observation angle. Here commas denote derivatives with respect to directions perpendicular to the line of sight. The trace of the distortion tensor represents the convergence
\begin{equation}
\kappa(\hat{\mathbf n})=\frac{1}{2}\left(\phi_{,11}(\hat{\mathbf n})+\phi_{,22}(\hat{\mathbf n})\right)
\end{equation}
and, defining $\gamma_1(\hat{\mathbf n})=\frac{1}{2}\left(\phi_{,11}(\hat{\mathbf n})-\phi_{,22}(\hat{\mathbf n})\right)$ and $\gamma_2(\hat{\mathbf n})=2\phi_{,12}(\hat{\mathbf n})$, the linear combination
\begin{equation}
\gamma(\hat{\mathbf n})=\gamma_1(\hat{\mathbf n})+i\gamma_2(\hat{\mathbf n})
\end{equation}
is the differential stretching, or shear.

\subsection{Power spectra}\label{powerspectra}

The study of the weak lensing effects can be made by analysing the quantities introduced above. Now we will present the power spectra of the projected potential, convergence and shear, on the all sky and in the flat--sky approximation.

\subsubsection{Weak lensing power spectra on the all sky}

By decomposing the projected potential \eqref{phi} in spherical harmonics $Y^m_l$, we obtain
\begin{equation}
\phi_{lm}(\mathbf x)=4\pi i^l\int\!\!\frac{\de^3k}{{(2\pi)}^3}\mathbf{\Phi}_k(0)I^\phi_l(k){Y^m_l}^\ast(\hat{\mathbf k})
\end{equation}
with
\begin{equation}
I^\phi_l(k)=\int\!\!\de\chi\,\frac{\mathbf{\Phi}_k(\chi)}{\mathbf{\Phi}_k(0)}W(\chi)j_l(k\chi)
\end{equation}
where $j_l(k\chi)$ are the spherical Bessel's functions of order $l$. Remembering the definition of variance
\begin{equation}
\langle\phi^\ast_{lm}\phi_{l'm'}\rangle=\delta_{ll'}\delta_{mm'}C^{\phi\phi}_l,
\end{equation}
the power spectrum takes the form
\begin{equation}
C^{\phi\phi}_l=\frac{2}{\pi}\int\!\!\de k\,k^2{[I^\phi_l(k)]}^2P^\mathbf{\Phi}(k,\chi)
\end{equation}
where $P^\mathbf{\Phi}(k,\chi)$ is the 3D power spectrum of $\mathbf{\Phi}(\hat{\mathbf n},\chi)$. The relations between the power spectrum of the projected potential $\phi(\hat{\mathbf n})$ and the most important quantities of the distorsion tensor, the convergence $\kappa(\hat{\mathbf n})$ and the shear $\gamma(\hat{\mathbf n})$, can be derived via the formalism of the spin weighted spherical harmonics \citep{Stebbins:1996wx,Hu:2000ee,Castro:2005bg}; we obtain
\begin{align}
\kappa(\hat{\mathbf n})&=-\sum_{l=1}^\infty\sum_{m=-l}^l\,\frac{l}{2}(l+1)\phi_{lm}Y^m_l(\hat{\mathbf n}),\\
\gamma(\hat{\mathbf n})&=\sum_{l=1}^\infty\sum_{m=-l}^l\,\frac{1}{4}\sqrt{\frac{(l+2)!}{(l-2)!}}\phi_{lm}\phantom i_{\pm2}Y^m_l(\hat{\mathbf n}).\label{Cl}
\end{align}
where $\phantom i_{\pm s}Y^m_l(\hat{\mathbf n})$ are the spherical harmonics $Y^m_l(\hat{\mathbf n})$ of spin weight $\pm s$. From these relations the convergence and shear power spectra are
\begin{align}
C^{\kappa\kappa}_l&=\frac{l^2}{4}{(l+1)}^2C^{\phi\phi}_l,\label{Clkk}\\
C^{\gamma\gamma}_l&=\frac{1}{4}\frac{(l+2)!}{(l-2)!}C^{\phi\phi}_l.\label{Clgg}
\end{align}

\subsubsection{Weak lensing power spectra in the flat--sky approximation}

In the flat--sky approach we expand the projected potential $\phi(\hat{\mathbf n})$ in its Fourier modes
\begin{equation}
\phi(\hat{\mathbf n})=\int\!\!\frac{\de^2l}{{(2\pi)}^2}\,\phi(\mathbf l)e^{i\mathbf l\cdot\hat{\mathbf n}}.
\end{equation}
The power spectrum is defined as the Fourier transform of the correlation function:
\begin{equation}
\langle\phi^\ast(\mathbf l)\phi(\mathbf l')\rangle={(2\pi)}^2\delta_D(\mathbf l-\mathbf l')C^{\phi\phi}(l)
\end{equation}
where $\delta_D(\mathbf l-\mathbf l')$ is Dirac's $\delta$ function. We thus have \citep{Kaiser:1991qi}
\begin{equation}
C^{\phi\phi}(l)=\int_0^\infty\!\!\de\chi\,\frac{W^2(\chi)}{\chi^6}P^\mathbf{\Phi}\left(\frac{l}{\chi},\chi\right)\label{C(l)}
\end{equation}
where we introduced Limber's approximation, in which the only Fourier modes that contribute to the integral are those with $l\gg k\chi$.

To write the convergence and shear power spectra in the flat--sky approximation, we can follow two different routes: we can either remember that, due to Limber's approximation, the flat--sky power spectrum is equivalent to the all-sky one at large enough $l$, so we can take the limit of Eqs.~(\ref{Clkk}) and~(\ref{Clgg}) for large $l$, or we can do all the calculations expressing the distortion tensor in Fourier's modes. It is very easy to verify that the two approaches are equivalent, because the Fourier transform of the distortion tensor simply acts by multipling the projected potential by a factor $l$ for each derivative.

\section{Results}\label{results}

In this section we present the weak lensing power spectra for the CMB and background galaxies.

The cosmological parameters we use are: $h=0.72$, $\Omega_m=0.26$ and $\Omega_\Lambda=0.74$. We use the BBKS transfer function \citep{Bardeen:1985tr}, and the WMAP5 normalization $\sigma_8=0.796$ \citep{Dunkley:2008ie}. We also compute the errors on the $\Lambda$CDM $C^{\kappa\kappa}(l)$ signal \citep{Kaiser:1991qi,Kaiser:1996tp,Hu:1998az} as
\begin{equation}
\Delta C^{\kappa\kappa}(l)=\sqrt{\frac{2}{(2l+1)f_\mathrm{sky}}}\left(C^{\kappa\kappa}(l)+4\left\langle{\gamma_\mathrm{int}}^2\right\rangle/\bar n\right),\label{errors}
\end{equation}
where $\bar n$ is the number of galaxies per sq degree, $f_\mathrm{sky}={\Theta_\mathrm{deg}}^2\pi/129600$ is the fraction of the sky covered by a survey of dimension $\Theta_\mathrm{deg}$ (in degrees) and ${\left\langle{\gamma_\mathrm{int}}^2\right\rangle}^{0.5}\approx0.4$ is the typical intrinsic galaxy ellipticity. We apply Eq.~\eqref{errors} to a $20,000$ sq degree survey like those expected to be performed by EUCLID\footnote{\texttt{http://www.dune-mission.net}} \citep{Refregier:2008nq,Refregier:2008js} or Pan--STARRS\footnote{\texttt{http://Pan-STARRS.ifa.hawaii.edu}} \citep{2002SPIE.4836..154K,2002AAS...20112207K}. In Fig.~\ref{cl-cmb}, \ref{cl-bg_kaiser} and \ref{cl-bg_z0=1} we show error boxes with $\bar n=4.7\cdot10^8\,\mathrm{deg}^{-2}$.

In UDM models, the background evolution of the Universe is the same as in $\Lambda$CDM, while the evolution of the gravitational potential and the growth of LSS suffer the non negligible sound speed, that increases with time. The discriminant is the effective Jeans' length of the gravitational potential $\lambda_J(a)$. The Newtonian potential in UDM models behaves like in the $\Lambda$CDM model at scales much larger than $\lambda_J(a)$, while at smaller scales it starts to decay and oscillate. Weak lensing observables, like cosmic convergence or shear, are an integral over the line of sight, hence they do not show directly these oscillations. However, high values of $l$, that in
Limber's approximation play the role of the multipoles of Legendre's polynomials, correspond to small scales, and thus cosmic convergence at high $l$'s must show the decay of the deflecting potential.

Currently there is no linear-to-non--linear mapping in UDM models, thus our analysis is made according to the linear theory of perturbations, and what we calculate is not correct for any $l$. The multipole is related to the scale $k$ by the direct proportionality $k=l/\chi(z)$. We estimate the window of multipoles $l$ of validity of our approximations in the following way: the lower limit is $l\simeq10$, due to Limber's approximation, but the upper one is floating, because at higher multipoles non--linear effects become more important. Weak lensing power spectra are made by integrating over the line-of-sight, thus over the wide range of the angular comoving distance $\chi(z)$. When we deal with high multipoles, wave numbers $k>k_\mathrm{nl}\simeq0.2\,h\,\mathrm{Mpc}^{-1}$ will appear in the integration. It is known that the linear theory $P^\mathbf{\Phi}(k>k_\mathrm{nl})$ underestimates the non--linear $P^\mathbf{\Phi}_\mathrm{nl}(k)$. To estimate the upper limit of validity of our results we perform the $\de\chi$ integration for $C^{\phi\phi}(l)$ using only the linear power spectrum for the gravitational potential, obviously obtaining a lensing signal lower than the correct one. Then, we perform the same integral by imposing a lower cut-off at $\chi_\mathrm{nl}\equiv l/k_\mathrm{nl}$. What we get in this way is a weak lensing power spectrum much more suppressed at large $l$'s than the one obtained in the former integration. With these two quantities, $C^{\phi\phi}_{\chi_\mathrm{min}=0}(l)$ and $C^{\phi\phi}_{\chi_\mathrm{min}=\chi_\mathrm{nl}}$, we fix an arbitrary threshold $\varepsilon$ such that
\begin{equation}
\varepsilon>\left|\frac{C^{\phi\phi}_{\chi_\mathrm{min}=0}(l)-C^{\phi\phi}_{\chi_\mathrm{min}=\chi_\mathrm{nl}}(l)}{C^{\phi\phi}_{\chi_\mathrm{min}=0}(l)}\right|;\label{epsilon}
\end{equation}
$\varepsilon$ enables us to estimate $l_\varepsilon$: for $l<l_\varepsilon$ most of the signal comes from the linear regime, and for $l>l_\varepsilon$ our ignorance on non--linear effects is too high that the real power spectrum could be substantially different from what we obtain. We present two thresholds $\varepsilon=10\%,20\%$ corresponding to the dashed vertical lines in Figs.~\ref{cl-cmb}, \ref{cl-bg_kaiser} and \ref{cl-bg_z0=1}.

We confirm the validity of our Eq.~\eqref{epsilon} by computing the non--linear weak lensing power spectra in $\Lambda$CDM with the \citet{Peacock:1996ci} non--linear modification to the linear matter power spectrum. \citet[Fig. 1]{Cooray:2003ar}, compared weak lensing linear and non--linear power spectra for sources at different redshifts; he showed that the higher is the value of the redshift $z_p$ where the distribution of the source redshifts peaks, the wider is the range of multipoles where linear theory is valid. This means that the multipole $l_\varepsilon$ at which non--linear effects become dominant is greater for sources located at higher redshift. With our estimation, we find roughly the same results.

\subsection{CMB}\label{cmb}

In this case, the source is the last scattering surface at $z_\mathrm{rec}\simeq10^3$, and its distribution is
\begin{equation}
n_\mathrm{CMB}\left[\chi(z)\right]=\delta_D\left[\chi(z)-\chi(z_\mathrm{rec})\right].
\end{equation}

\begin{figure}
\centering
\input{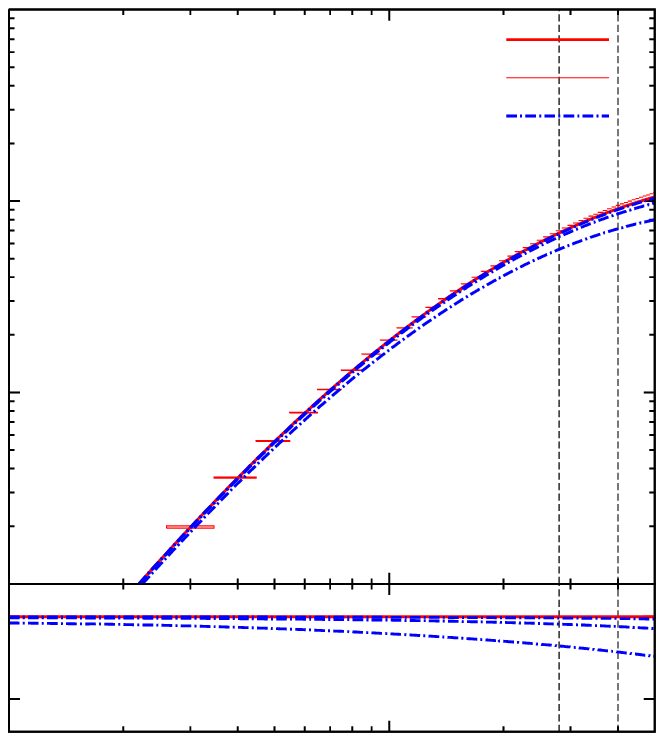}
\caption{Upper panel: power spectra $l(l+1)C^{\kappa\kappa}(l)/(2\pi)$ of CMB photons for $\Lambda$CDM (solid) and UDM (dot-dashed), for ${c_\infty}^2=10^{-6},10^{-5},10^{-4}$ from top to bottom. For CMB light, the linear (thin line) and non--linear (thick line) power spectra are almost on top of each other. Lower panel: ratio $C^{\kappa\kappa}(l)/C_\mathrm{lin.\,\Lambda CDM}^{\kappa\kappa}(l)$ for $\Lambda$CDM and UDM, one for each value of ${c_\infty}^2$ plotted in the upper panel. The vertical lines denote $l_\varepsilon$, the upper limit we estimate for the reliability of our results in the linear approximation, with the two assumptions for the threshold $\varepsilon=10\%,20\%$. The boxes show $1\sigma$ errors on the $\Lambda$CDM power spectrum according to a EUCLID--like $20,000$ sq degree survey.}\label{cl-cmb}
\end{figure}
In Fig.~\ref{cl-cmb}, the upper panel shows the weak lensing power spectra $l(l+1)C^{\kappa\kappa}(l)/(2\pi)$ of CMB light for $\Lambda$CDM and UDM. For the $\Lambda$CDM we show both the linear (thin line) and non--linear (thick line) power spectra. For UDM, we present three curves, obtained for ${c_\infty}^2=10^{-6},10^{-5},10^{-4}$. In the lower panel the UDM curves of the upper panel are divided by the linear convergence power spectrum of $\Lambda$CDM. As we can see, for small values of the sound speed (${c_\infty}^2=10^{-6}$), we cannot distinguish the convergence of CMB photons in UDM models from the standard $\Lambda$CDM behaviour. By increasing $c_\infty$, while at large scales the behaviour is similar, at small enough scales $C^{\kappa\kappa}_\mathrm{UDM}(l)$ is clearly suppressed. The errors for a EUCLID--like survey show that if the sound speed is greater than $10^{-2}$ we are able to discriminate UDM from $\Lambda$CDM even at scales $l\lesssim l_\varepsilon$.

\subsection{Background galaxies}\label{backgroundgalaxies}

In this case, the sources are spread over different redshifts, and the distribution can be assumed to be \citep{Kaiser:1991qi}
\begin{equation}
n_g\left[\chi(z)\right]=\frac{\beta z^\alpha}{{z_0}^{\alpha+1}}\frac{e^{{-\left(\frac{z}{z_0}\right)}^\beta}}{\Gamma\left(\frac{\alpha+1}{\beta}\right)}\frac{\de z}{\de\chi},\label{source-distribution}
\end{equation}
that peaks at redshift $z_p\equiv z_p(\alpha,\beta,z_0)$, where $\alpha$, $\beta$ and $z_0$ are free parameters.

\begin{figure}
\centering
\input{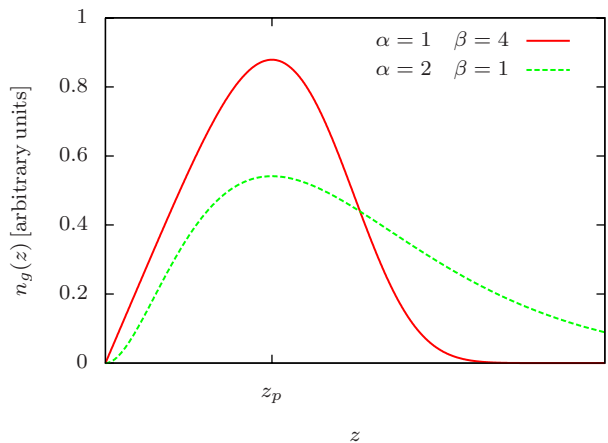}
\caption{The redshift distributions $n_g(z)$ of galaxies (Eq.~\ref{source-distribution}). We use here: $\alpha=1$ and $\beta=4$ of \protect\citet{Kaiser:1991qi} (solid) and $\alpha=2$ and $\beta=1$ of \protect\citet{Wittman:2000tc} (dashed). The curves are presented with an arbitrary $z_p$ (by fixing it we obtain $z_0$) to show their relative shapes.}\label{filters}
\end{figure}
We want to see how the redshift distribution of the sources affects the power spectrum of weak lensing. To do so, we will use different functional forms for $n_g(z)$: in Fig.~\ref{filters} we show the redshift distributions we use in this paper. To study the background galaxy light we will use three different source distributions, two of them varying the parameters in Eq.~\eqref{source-distribution}, as suggested by \citet{Kaiser:1991qi} and \citet{Wittman:2000tc}, and one using a  Dirac's delta function at $z=z_p$ \citep{Hu:1998az}. We will compute the convergence power spectrum for $z_p=1,2,3$ \citep{Cooray:2003ar}.

\begin{figure}
\centering
\input{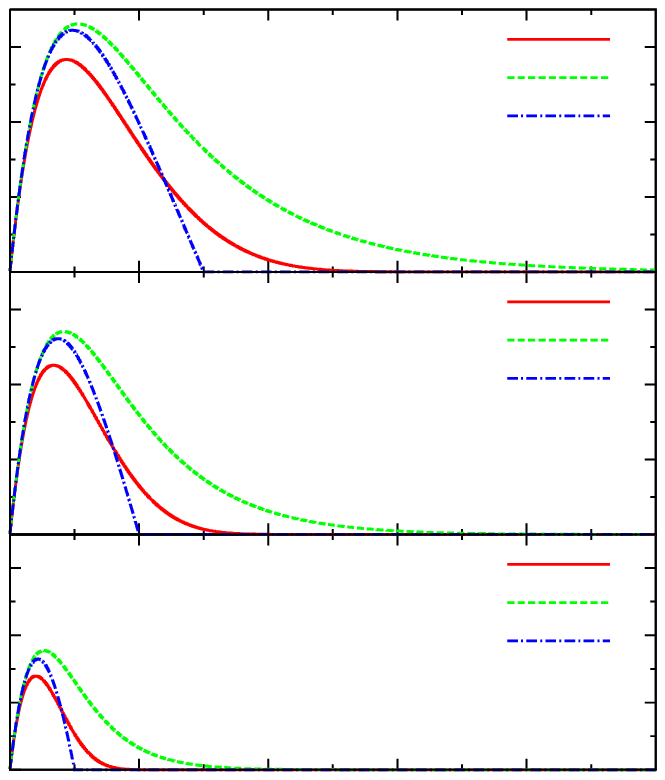}
\caption{Weight functions for background galaxies with $n_g(z)$ of Eq.~\eqref{source-distribution} with $\alpha=1$ and $\beta=4$ \protect\citep{Kaiser:1991qi} (solid), $\alpha=2$ and $\beta=1$ \protect\citep{Wittman:2000tc} (dashed) and with a Dirac's delta at $z=z_p$ (dot-dashed).
}\label{W_g}
\end{figure}

To better understand our results, it is useful to look at the weight functions $W_g(z)$ for background galaxies (Eq.~\ref{W(z)}). In Fig.~\ref{W_g} we present $|W_g(z)|$ for the different choices of the source redshift distributions and the three $z_p$'s we use in this work. Differences between convergence power spectra with the same redshift distribution of the sources and different $z_p$'s, or vice versa, are due to these shapes.  The height of the peak of $|W_g(z)|$ determines the order of magnitude of the weak lensing signal.

\begin{figure*}
\centering
\input{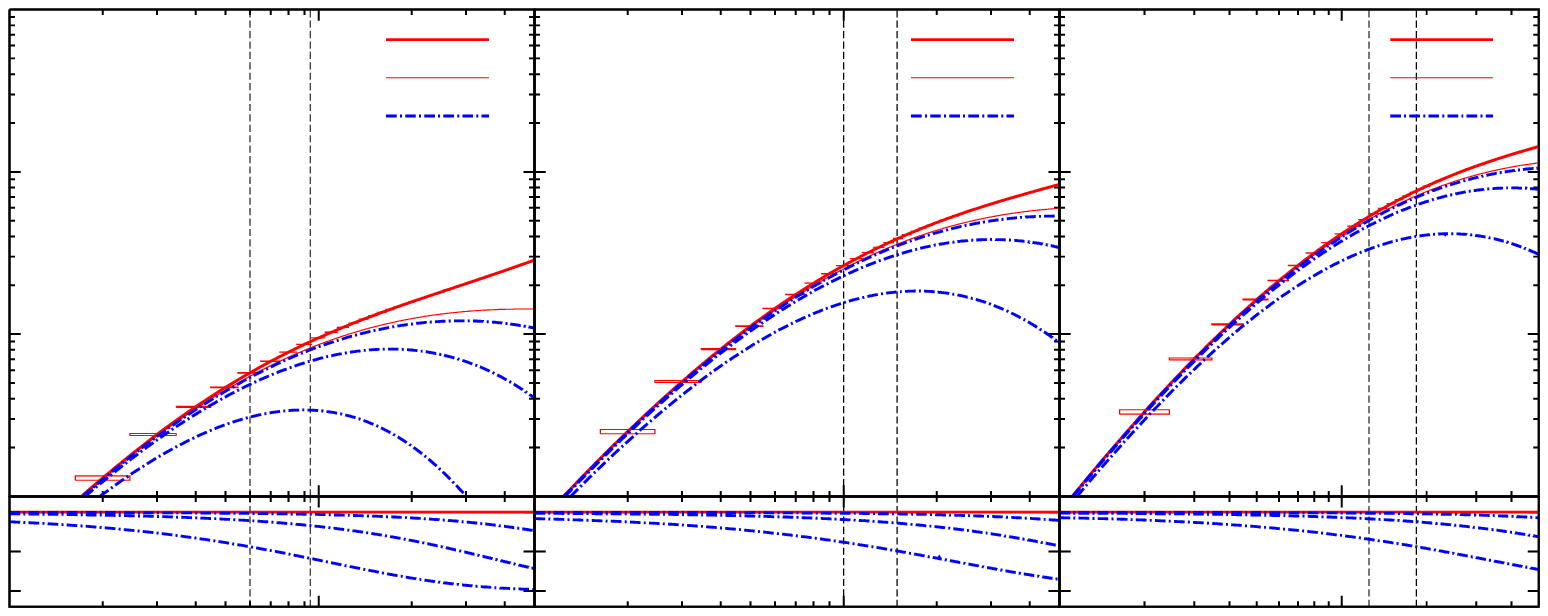}
\caption{Same as Fig.~\ref{cl-cmb} for background galaxies. The parameters of the source redshift ditribution are $\alpha=1$ and $\beta=4$ (the solid line in Fig.~\ref{filters}). In the left panel $z_p=1$, in the middle panel $z_p=2$ and in the right panel $z_p=3$.}\label{cl-bg_kaiser}
\input{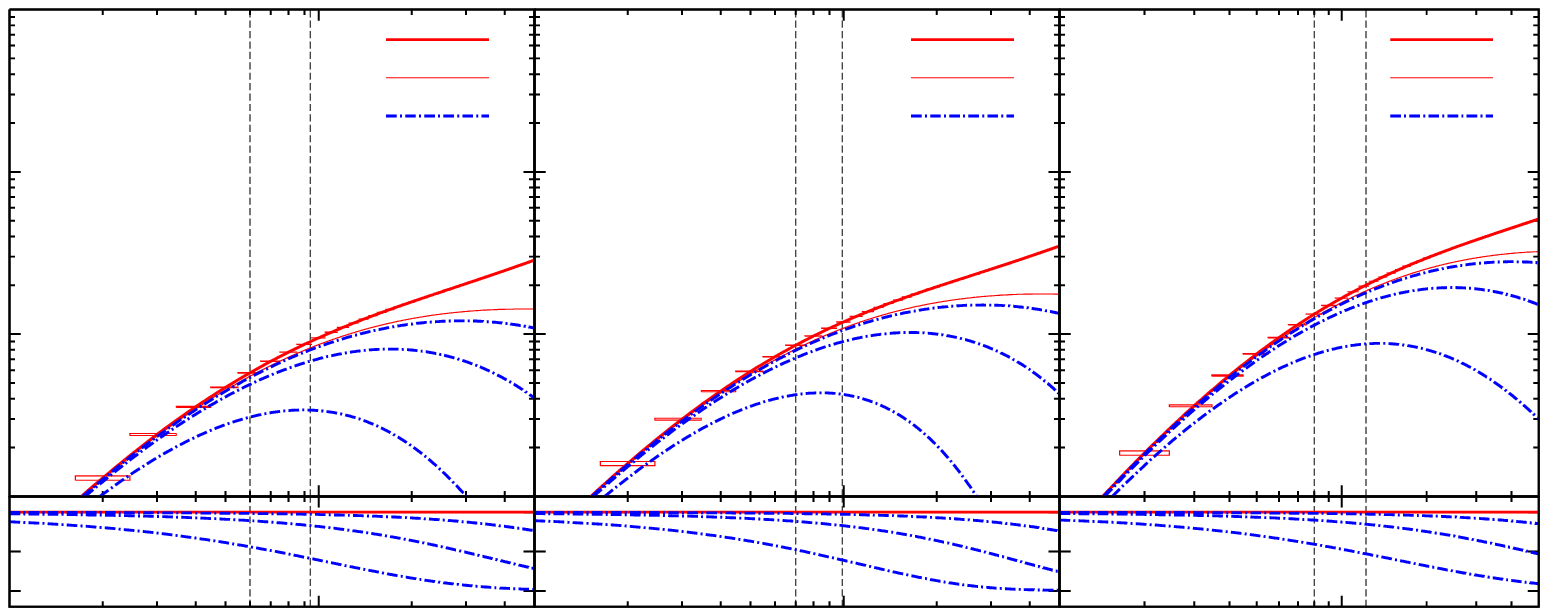}
\caption{Same as Fig.~\ref{cl-cmb} for background galaxies with different redshift distributions peaked at $z_p=1$. Left panel: $\alpha=1$ and $\beta=4$ (the solid line in Fig.~\ref{filters}); middle panel: $\alpha=2$ and $\beta=1$ (the dashed line in Fig.~\ref{filters}); right panel: $n_g(z)$ is Dirac's delta function at $z=z_p$.}\label{cl-bg_z0=1}
\end{figure*}
In Figs.~\ref{cl-bg_kaiser} and \ref{cl-bg_z0=1}, upper panels, we show the weak lensing power spectra $l(l+1)C^{\kappa\kappa}(l)/(2\pi)$ of background galaxy light for $\Lambda$CDM and UDM. In particular, we present three peak redshifts $z_p$ for a single distribution of the sources (Fig.~\ref{cl-bg_kaiser}) and three redshifts distribution of the sources $n_g(z)$ at a single peak redshift (Fig.~\ref{cl-bg_z0=1}). As for the CMB case, for $\Lambda$CDM we show both the linear (thin line) and non--linear (thick line) power spectrum, and for UDM we present three curves, obtained for ${c_\infty}^2=10^{-6},10^{-5},10^{-4}$. In the lower panels the UDM curves of the upper panels are divided by the linear convergence power spectrum of $\Lambda$CDM. As we can see, for small sound speeds (${c_\infty}^2=10^{-6}$) and large angular scales ($l\lesssim100$), we cannot distinguish the convergence of background galaxy photons in UDM models from the standard $\Lambda$CDM behaviour. However, the agreement disappears at large $c_\infty$ and $l$'s. For background galaxy light UDM features are significant, and EUCLID will enable us to distinguish between UDM and $\Lambda$CDM if ${c_\infty}^2\gtrsim10^{-4}$, similarly to the CMB result. The difference between $\Lambda$CDM and UDM is larger with background galaxies than with CMB because the background galaxy signal is more sensitive to the sound speed. Moreover, the weak lensing signal decreases with decreasing redshift of the background galaxies.

The power spectrum $P^\mathbf{\Phi}(k,z)$ of the gravitational potential, through Poisson's equation, encodes the distribution of overdensities, thus the structures crossed by the photons. The weight function $W(\chi)$ is a filter that selects mostly signals emitted at $z\lesssim z_p$, following the information of $n(z)$. Consequently, we can notice in Figs.~\ref{cl-cmb}, \ref{cl-bg_kaiser} and \ref{cl-bg_z0=1}, for different values of the speed of sound, how the weak lensing signal in UDM models is sensitive to the choice of $z_p$. At the same time, $c_s$ has to be very small to let the scalar field cluster and form the LSS we observe today; on the contrary in the past, at high enough redshift, the gravitational potential is similar to that predicted by $\Lambda$CDM. However, sources at lower $z_p$ emit light that strongly feels the decay and the oscillations of the Newtonian potential, because this potential is sensitive to the sound horizon $\lambda_J\equiv\lambda_J(z,c_\infty)$, that increases with time (see Fig.~\ref{lambda_J}), and to the presence of an effective $\Omega_\Lambda$, that plays the role of DE. It is easy to see that, for small $z_p$, the differences between UDM models with different $c_\infty$ and between UDM and $\Lambda$CDM are very pronounced even at large angular scales, while for CMB the power spectrum is less sensitive to the sound speed. Finally we observe that, at fixed $z_p$ and $l$, if the peak of $|W(z)|$ is as close to zero as possible (Fig.~\ref{cl-bg_kaiser}), we find a higher dependence of $C^{\phi\phi}(l)$ on the sound speed $c_\infty$. This can help to choose an appropriate redshift distribution $n_g(z)$ to constrain $c_\infty$ from weak lensing.

\subsection{Uncertanties of cosmological parameters on the $\Lambda$CDM weak lensing signal}

\begin{figure*}
\centering
\input{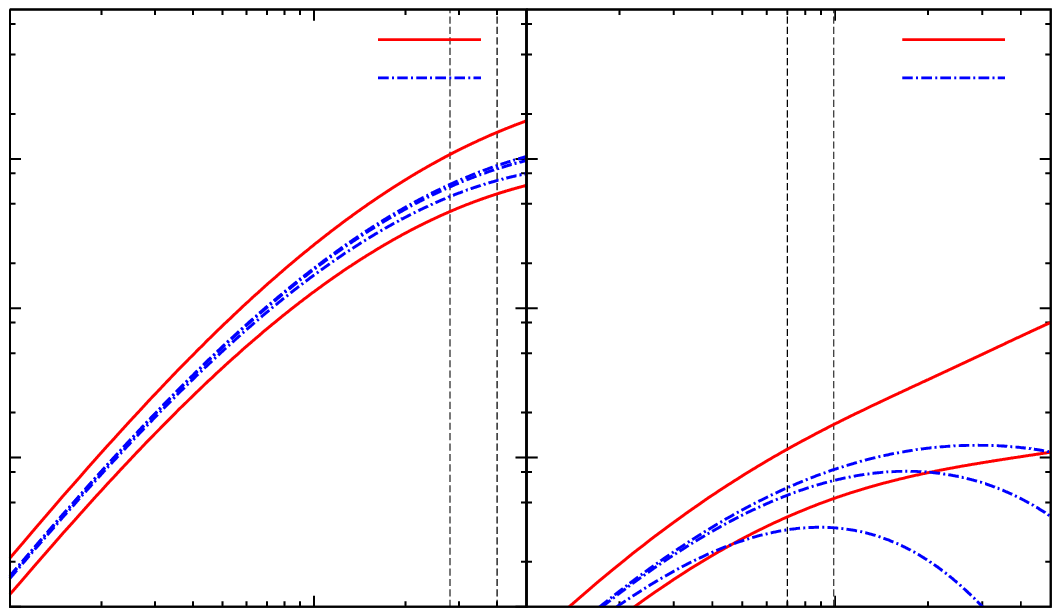}
\caption{Role of the $68\%$ uncertainties of $\Omega_m$ and $\sigma_8$ on the $\Lambda$CDM signal. The top and bottom solid lines represent the upper and lower limits of the non--linear convergence power spectrum in the $\Lambda$CDM model due to these uncertenties; the dot-dashed curves are UDM power spectra with ${c_\infty}^2=10^{-6},10^{-5},10^{-4}$, from top to bottom. In the left panel, the source is the CMB, in the right panel the sources are background galaxies peaked at $z_p=1$ and distributed over a range of angular diameter distances according to Eq.~\eqref{source-distribution} with $\alpha=1$ and $\beta=4$ \protect\citep{Kaiser:1991qi}.}\label{sigma_8-Omega_m}
\end{figure*}

In the $\Lambda$CDM model, we compute the convergence power spectra for different values of $\Omega_m$ and $\sigma_8$ within the $68\%$ region of their uncertainties, as derived by \citet{Dunkley:2008ie} with WMAP5 data (their Fig.~3). In Fig.~\ref{sigma_8-Omega_m} (left panel) the top and bottom solid lines represent the upper and lower limits of the power spectra due to these uncertainties. We notice that the CMB is not a good source to constrain UDM models with weak lensing because, even if the errors on the signal are very small, as we show in \S \ref{cmb}, the UDM convergence power spectrum lies within the $\Lambda$CDM strip. On the contrary, if we use the convergence power spectrum of background galaxies (right panel), we are in principle still able to distinguish between the signals of $\Lambda$CDM and UDM models.

Finally, we find that it is not possible to reproduce a UDM power spectrum with a fixed $c_\infty\neq0$ by properly choosing $\Omega_m$ and $\sigma_8$ in the $\Lambda$CDM model, even if the former lies within the uncertainty of the latter. In fact, the dependence of the Newtonian potential in UDM models on scale and redshift is not factorizable in a scale-dependent transfer function $T(k)$ and a redshift-dependent growth factor $D(z)$, but the two dependences are linked together by the sound horizon $\lambda_J(z)$.

\section{Conclusions}\label{conclusions}

In this work, we investigate weak gravitational lensing in models of unified DM and DE. We focus on those UDM models which are able to reproduce the same background expansion history as in the $\Lambda$CDM model \citep{Bertacca:2007ux,Bertacca:2008uf}. When we switch cosmological perturbations on, differences arise. In $\Lambda$CDM the Universe is filled with a perfect fluid of radiation, cold DM and baryons, and the vacuum energy of the cosmological constant. In UDM, beyond standard matter and radiation, there is only one exotic component, a classical scalar field with a non-canonical kinetic term, that during structure formation behaves like matter, while at the present time contributes to the total energy density of the Universe like a cosmological constant.

A general severe problem of many UDM models is that their large effective speed of sound causes a strong time evolution of the gravitational potential and generates an integrated Sachs-Wolfe effect much larger than current observational limits. Recently, \citet{Bertacca:2008uf} outlined a technique to reconstruct UDM models such that the sound speed is small enough that these problems are removed and the scalar field can
cluster.

Here, we show the lensing signal in linear theory as produced in $\Lambda$CDM and UDM; as sources, we consider the CMB and background galaxies, with different values of the peak and different shapes of their redshift distribution. For sound speed lower than $c_\infty=10^{-3}$, in the window of multipoles $l\gtrsim10$ (Limber's approximation) and $l<l_\varepsilon$ (where our ignorance on non--linear effects due to small scales dynamics become relevant), the power spectra of the cosmic convergence (or shear) in the flat--sky approximation in UDM and $\Lambda$CDM are similar. When the Jeans' length $\lambda_J(a)$ increases, the Newtonian potential starts to decay earlier in time (at a fixed scale), or at greater scales (at a fixed epoch). This behaviour reflects on weak lensing by suppressing the convergence power spectra at high multipoles. We find that, for values of the sound speed between $c_\infty=10^{-3}$ and $c_\infty=10^{-2}$, UDM models are still comparable with $\Lambda$CDM, while for higher values of $c_\infty$ these models are ruled out because of the inhibition of structure formation. Moreover, we find that the dependence of the UDM weak lensing signal on the sound speed $c_\infty$ increases with decreasing redshift of the sources. We also show the errors for the fiducial $\Lambda$CDM signal for wide--field surveys like EUCLID OR Pan--STARRS, and we find that we are in principle able to distinguish $\Lambda$CDM from UDM models when $c_\infty\gtrsim10^{-2}$.

\subsection*{Acknowledgments}

We thank the referee Lauro Moscardini for a careful reading of the manuscript and for his comments, that helped to improve substantially the presentation of our results. Most of this work was completed when DB was supported by the PRIN2006 grant ``Costituenti fondamentali dell'Universo'' of the Italian Ministry of University and Scientific Research at the Department of Theoretical Physics and the Department of General Physics ``Amedeo Avogadro'' of the University of Torino. SC and AD also acknowledge partial support from this grant. Partial support from the INFN grant PD51 is also gratefully acknowledged. This research has been partially supported by ASI contract I/016/07/0 ``COFIS''.

\bibliographystyle{mn2e3}
\bibliography{WlsiUDMm_bibliography}

\end{document}